\pdfoutput=1
\title{Visualizing the Structure of Large Trees}
\author{
        Burcu Ayd{\i}n \footnote{Partially supported by NSF grants DMS-0606577 and DMS-0854908, and NIH Grant RFA-ES-04-008.}
        \and
        G\'{a}bor Pataki
        \and
        Haonan Wang\footnote{Partially supported by NSF grants DMS-0706761 and DMS-0854903.}
        \and
        Alim Ladha
        \and
        Elizabeth Bullitt\footnote{Partially supported by NIH grants R01EB000219-NIH-NIBIB and R01 CA124608-NIH-NCI.}
        \and
        J.S. Marron \footnote{Partially supported by NSF grants DMS-0606577 and DMS-0854908, and NIH Grant RFA-ES-04-008.}
}

\date{\today}

\documentclass[12pt]{article}
\usepackage{graphicx}

\begin{document}
\maketitle

\begin{abstract}
This study introduces a new method of visualizing complex tree structured objects.
The usefulness of this method is illustrated in the context of detecting unexpected features in a data set of very large trees.
The major contribution is a novel two-dimensional graphical representation of each tree, with a covariate coded by color.

The motivating data set contains three dimensional representations of brain artery systems of $105$ subjects. Due to inaccuracies inherent
in the medical imaging techniques, issues with the reconstruction algorithms and inconsistencies introduced by manual adjustment, various discrepancies are present in the data. The proposed representation enables quick visual detection of the most common discrepancies. For our driving example, this tool led to the modification of $10\%$ of the artery trees and deletion of $6.7\%$.

The benefits of our cleaning method are demonstrated through a statistical hypothesis test on the effects of aging on vessel structure. The data cleaning resulted in improved significance levels.

\end{abstract}


\section{Introduction}\label{Introduction}

Most real life data sets contain a variety of challenging features, which can be noise artifacts or other kinds of discrepancies. Elimination of these artifacts may result in sharper statistical results. Relationships that were previously obscured may become more clear.

Our motivation comes from a data set of brain artery systems of $105$ subjects collected by the CASILAB (casilab.med.unc.edu). The extraction of this data set from raw Magnetic Resonance 
Angiography (MRA) images are summarized in Section \ref{Data}, and further details can be found in \cite{bullitt2002}.

\begin{figure}
[ptb]
\begin{center}
\includegraphics[
natheight=231.250000pt,
natwidth=330.750000pt,
height=163.875pt,
width=233.5pt
]%
{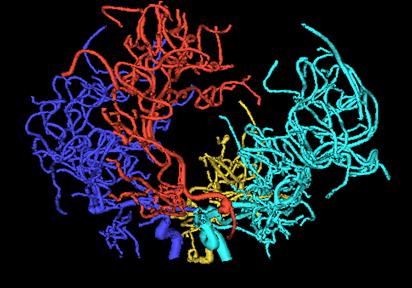}%
\includegraphics[
natheight=408.500000pt,
natwidth=395.125000pt,
height=206.25pt,
width=199.5625pt
]%
{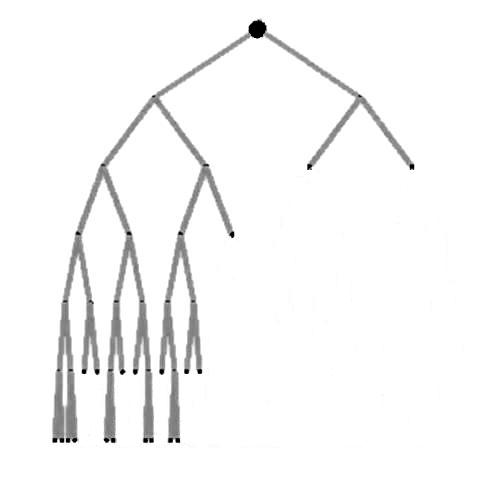}%
\caption{On the left: Reconstructed set of trees of brain arteries. The colors indicate regions of the brain: Gold (back),
Right (blue), Front (red), Left (cyan). On the right: Binary tree obtained from the gold (back tree) of the same subject. Only branching information is retained.}
\label{Fig1}%
\end{center}
\end{figure}

An earlier version of this data set consisting of $73$ data points (\texttt{Data Set $1$}) was used to statistically analyze the effect of aging in brain vessel structure in \cite{wang2007} and \cite{Aydin2009}.
In the latter paper, the rich three dimensional structure of the vessel systems are summarized by binary trees which only keep connectivity information. 
The aim was to strip the features other than branching from the data and obtain a simplified representation to study the effect of aging on the branching structure of the vessels. 
Figure \ref{Fig1} shows the $3-D$ image of a brain vessel system and one of the binary trees extracted from it.

Since the original analysis, $34$ more subjects were added to the study and two low quality cases were deleted. Careful anatomical examination of each data tree revealed some errors in the flow direction of vessels and revealed that starting points were arbitrary depending on the head position in the scanner.
The $3$ dimensional trees  (as in the left panel of Figure \ref{Fig1}) in \texttt{Data Set $1$} were constructed by an automatic vessel connection
algorithm (see \cite{bullitt2002} for details) which occasionally resulted in anatomically incorrect connections. These issues were addressed,
in a painstaking case by case fashion, by manual reconnection of tree components. The head position problem was addressed by starting each tree
at the \emph{circle of Willis} (well known, common component of human brains). When there were more than four trees, trees were combined to
result in the best approximation of front, left, right, back flow systems. The resulting data set is called \texttt{Data Set $2$}. As shown
in \cite{Aydin2009}, the same statistical methods applied to this new data set reveal  a remarkable improvement in significance levels. This demonstrates the value of improved data quality.


However, even after this cleaning process, many discrepancies are known to remain in the data set. The details of $3$ separate problems that we found in the data are explained in Section \ref{Identification}. Some of these discrepancies are fundamental in a way that they may change the observed structure of the vessel system. Elimination of as many of these problems as possible will improve the quality of a statistical analysis.

The full $3$ dimensional tree structures, e.g. as shown in the left panel of Figure \ref{Fig1}, contain a large amount of visual information. This makes it hard to see and understand the relevant problems with the data. The right panel of Figure \ref{Fig1} shows a simplified topology only structure, which enables focusing on purely topological aspects. However, this representation only works to node level $10$ to $11$ where there is not space to display more nodes. A major contribution of this paper is, due to the graphical representation that enables focusing on the important aspects of each data tree.

One problem with the $3-D$ representation is that, the amount of details present in the set is very high. Examining this set thoroughly requires checking the linkage and position of each branch in each subject separately, a n extremely tedious job to process manually. It is important to reduce the level of complication without losing the aspects necessary to track down discrepancies.

Graphically representing each data tree can be a powerful method to both understand the data and discover any problems. Carefully designing a visual representation of the data provides efficient visual inspection of each of these instances and may enable these discoveries without any further diagnostics. The challenge is to clearly display the important aspects of the data and eliminate the details that will not help with the diagnostics.

In this paper we propose a visualization method that aims to eliminate enough details to give a clear picture, but also keep enough to enable the discovery of discrepancies with ease.

Effective ways of visualizing trees has been previously studied in the literature. One example is the extensive work on phylogenetic trees. \cite{letunic2006} and \cite{huson2007} are two recent studies on phylogenetic trees which include good literature surveys on the subject for the further interested reader.

\cite{nyugen2002} and \cite{nyugen2007} bring an interesting approach to visualizing large trees on $2$ dimensional space. They use a concept called \textit{enclosure} to partition the entire display space into a collection of local regions that are assigned to each node of a tree. Search trees generated by finite domain constraint programs have also been investigated in the literature. \cite{Simonis2000} did a visualization study of such trees, using a software tool they develop to debug and analyze such trees. Finally, a recently popular approach is \textit{tree-maps}. Tree-maps transform the traditional branched tree view into a rectangle divided into sub-rectangles, arranged according to some properties of interest of the nodes. \cite{shneiderman1998} proposed the tree-map idea, and numerous variations have been studied in the literature.

The organization of this paper is as follows: Section \ref{Data} gives a description of how the binary trees used for statistical analysis are produced from the $3-D$ representations as shown in Figure 1. Section \ref{Visualization} develops the details of the proposed visualization method. Section \ref{Identification} indicates some  problems that commonly arise within the data set. Section \ref{Solution} explains how the visualization method is used to identify these problems. The results of the cleaning process together with a comparative statistical analysis is also given in this section.

\section{Extraction of Binary Trees From the Raw Data}\label{Data}

Following \cite{Aydin2009}, the way the binary trees are extracted is as follows: For each of the instances (brain scans), the back, left, right and front regions are handled separately. Each of these subsystems usually consist of one main (root) vessel entering the brain from below, and splitting into smaller branches to feed that region of the brain. The portion of the root vessel until a branch splits off is taken as the root node, and the two vessels that take place after the split are the left and right children nodes of the root. The same procedure is applied at each juncture point. In the end, a binary tree is obtained, where each vessel trunk between two split points in the original structure corresponds to a node.

An issue is, whether a vessel splits into three or more branches at a single point. In this data set splitting into more than two branches is very rare, and these occurrences do not carry any important implications on structure. Therefore keeping the simple binary structure seems more important than capturing these rare occurrences. In the cases that this happens, one of the child vessels is arbitrarily selected as the first one to split off, and the binary tree is created accordingly.

In some of the instances, there can be two root vessels feeding one region. In these cases, for simplicity, one phantom root node is added and the roots of these two trees are connected to the phantom root node as children, so that a single binary tree is obtained for each instance in the sub-populations. These binary trees are called \emph{component trees}.

\section{Visualization}\label{Visualization}

\begin{figure}
[ptb]
\begin{center}
\includegraphics[
natheight=9.375400in,
natwidth=12.500000in,
height=4in,
width=5in
]%
{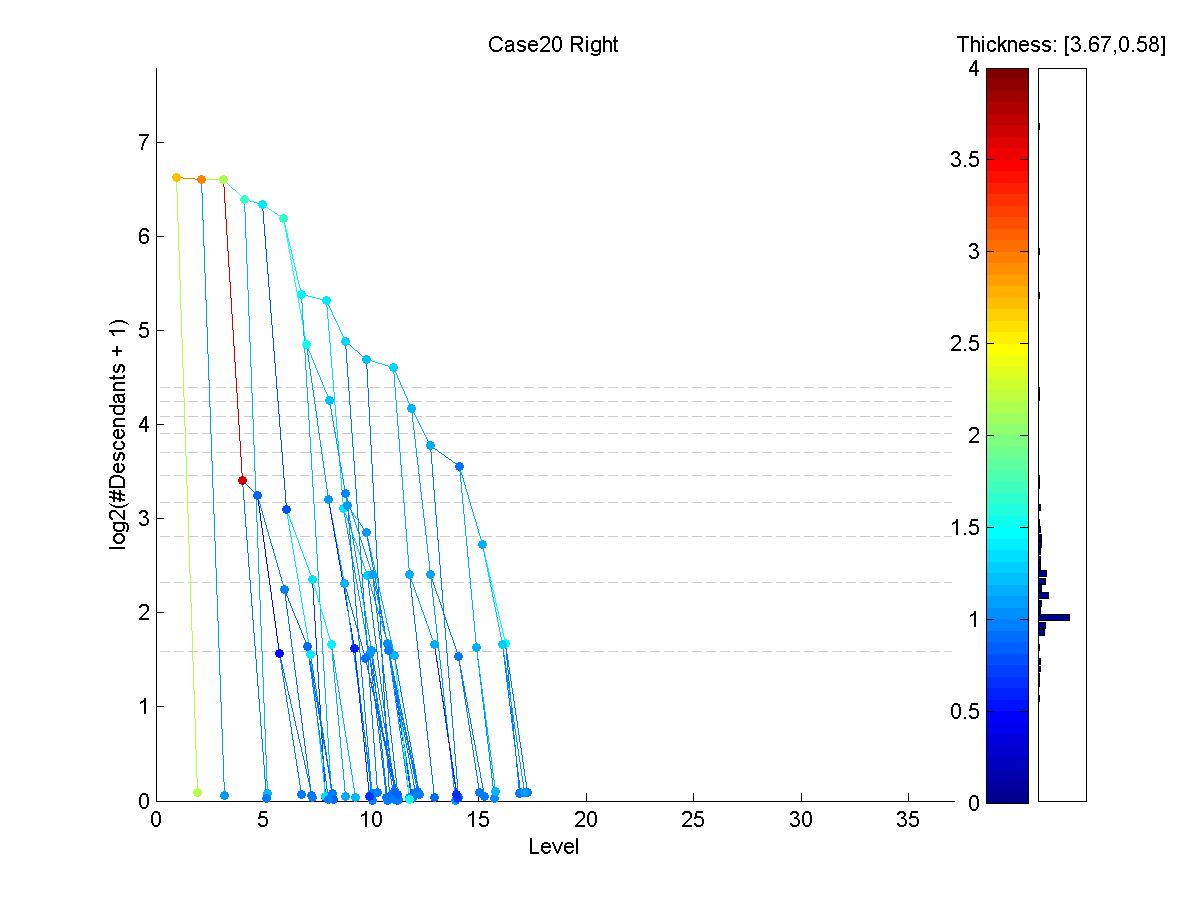}
\caption{A visual display of the right component tree of subject $60$, \texttt{Data Set $2$}.}
\label{20right}%
\end{center}
\end{figure}

An example of our visualization can be seen in Figure \ref{20right}. Each node is represented with a colored dot and each parent-child couple is connected with a line segment, revealing the connectivity. The y-coordinate of each node shows the logarithm (base $2$), for compact summarization, of the number of descendants. The x-axis shows the level of the node in the binary tree. Because of this axis selection, the visualization is called the \emph{D-L view} (Descendant-Level view).

In the lower levels, many nodes have the same coordinates (number of descendants and level). To avoid the over-plotting of these nodes, some jittering (adding small amounts of random displacement for improved visual separation) is applied at lower levels.

Because the thickness of the vessel segments is important to understand tree structures and potential errors, the median thickness is coded by color. In the brain vessel data set, this thickness rarely exceeds $4$ millimeters, so the range of thickness used is $[0,4]$. Thicker nodes are put in the top $4$mm bin. This range is linearly projected onto a color map which consists of $100$ shades, or bins, shown as the color bar on the right. The thicknesses close to $4 mm$ correspond to dark red shades. As vessels thin, the color follows the usual spectrum through yellow, green, to a dark shade of blue. The distribution of thicknesses is summarized in a bar chart, indicating the counts of the nodes that fall into the range of each bin, displayed on the right hand side of the figure. The numbers displayed on the right top corner show the thickness range of the nodes for that particular data tree, i.e. in Figure \ref{20right} the thinnest node has a thickness of $0.58 mm$.

\section{Identification Of Common Problems}\label{Identification}

Based on anatomical knowledge and experience, several types of discrepancies that may exist in the data have been identified. Note that some natural noise exists in this data along with tracking and labeling errors listed below. An important aim of the D-L view was to design a clear representation of the data such that these errors can be told apart from the natural noise. In Figure \ref{20right}, on level $4$, a vessel segment thicker than its parent exists, which is not a regular occurrence. However, the thickness difference of that segment with its parent is within measurement error range, so this instance is not flagged for checking.

The \texttt{Data Set $2$} is a result of a different, more anatomically based, case by case clean up process and the addition of $34$ more cases. That clean-up process aimed to correct the cut off point problems of the vessel systems, by manually going through each instance and examining them. This study is meant to develop a visual method to diagnose a wider range of problems without scrutinizing the raw data form as shown in Figure \ref{Fig1}, which is loaded with information and thus is hard to examine. The methods explained here are applied to \texttt{Data Set $2$}, as presented in Section \ref{Solution}, to obtain \texttt{Data Set $3$}.

We have identified the following major kinds of discrepancies that can be corrected through inspection:

\begin{itemize}
\item \textbf{Misconnections:} Some vessels that are not anatomically connected appear connected in the representation seen in left panel of Figure \ref{Fig1}, due to being close to each other relative to the accuracy of the MRA image slices. This error may result in misinterpretation of the blood flow of direction, and one of these vessels is seen as an extension of the other in the model. Normally, the vessel trunks that are connected to each other are expected to have similar thicknesses, and this thickness should generally decrease as one goes from root nodes to the leaves. A sudden jump in thickness is an indication of misconnection, i.e. when a thinner parent node has a much thicker sub-tree descending from it. Figure \ref{55back} shows an example of this situation. Notice that the red/yellow subtree starting from level $6$ is much thicker than its parent vessel.
\begin{figure}
[ptb]
\begin{center}
\includegraphics[
natheight=9.375400in,
natwidth=12.500000in,
height=4in,
width=5in
]%
{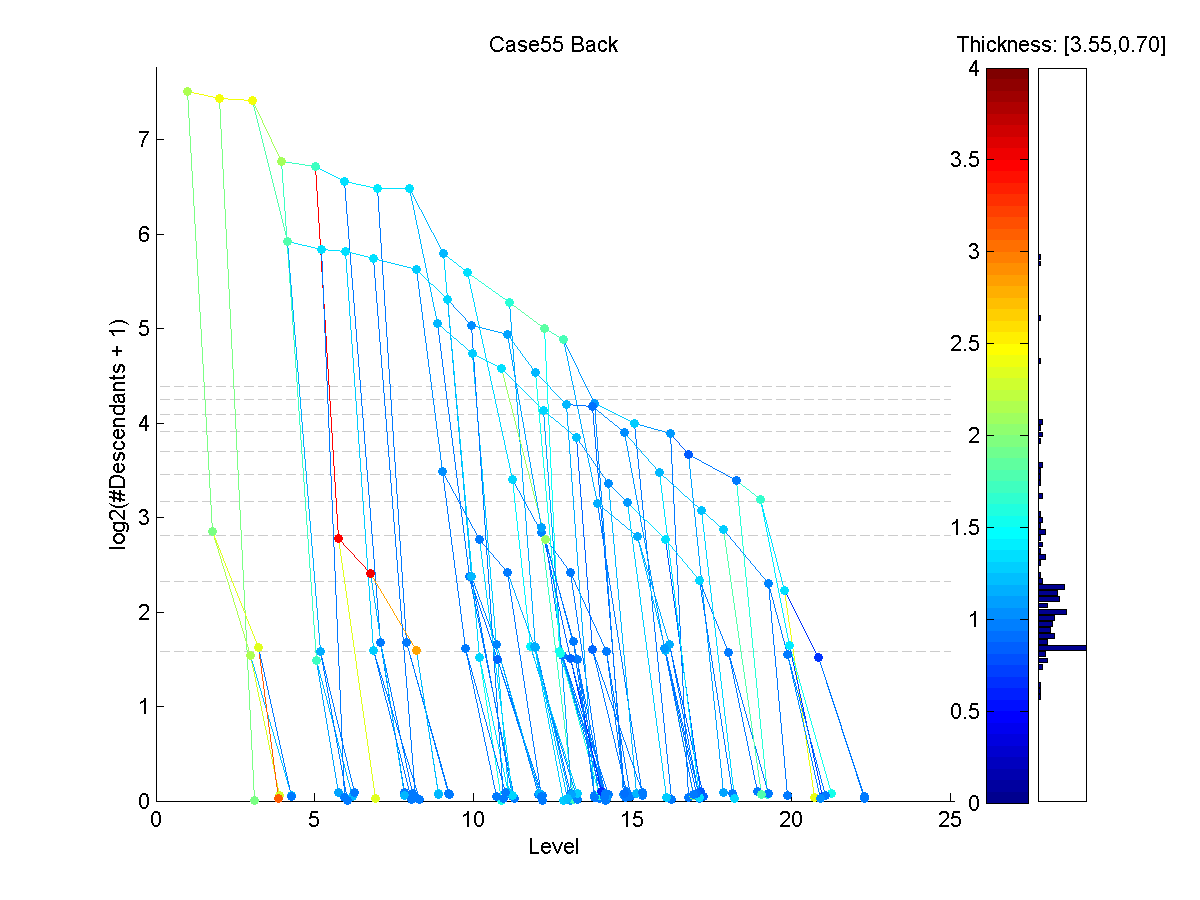}
\caption{The D-L view of the back component tree of subject $55$. Red/yellow subtree starting from level $6$ is a possible misconnection.}\label{55back}%
\end{center}
\end{figure}

\item \textbf{Starting Point Problems:} Determining the point where a vessel enters the brain highly depends on the cutoff point of the MRA images
(depending on head position in the scanner). In some cases, this cutoff point is mistakenly taken at a too low or too high level.
It is also possible to mark a child of a root as the root node. Figure \ref{28right} shows a tree with a possible starting point problem.
The initial series of thick red nodes suggest that the MRA starting point was taken too low.

\begin{figure}
[ptb]
\begin{center}
\includegraphics[
natheight=9.375400in,
natwidth=12.500000in,
height=4in,
width=5in
]%
{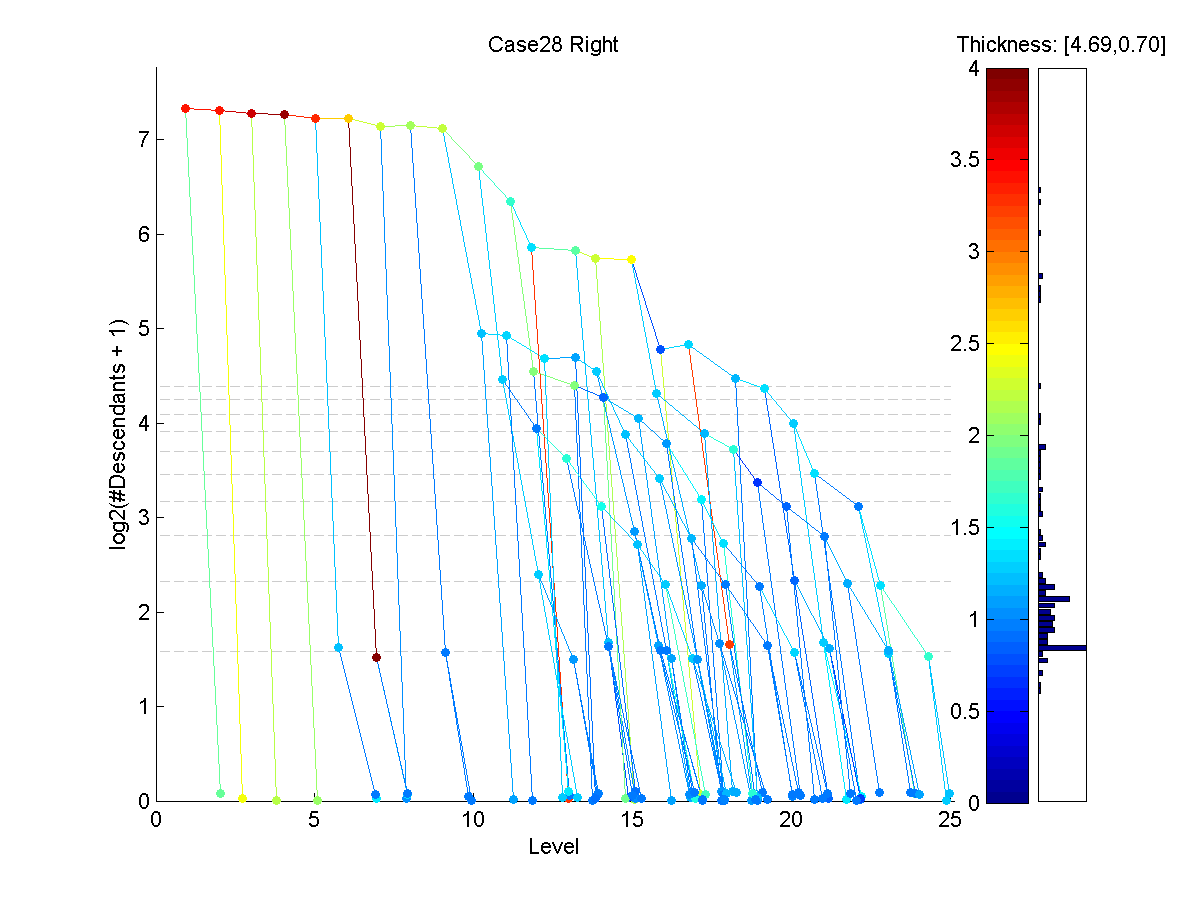}
\caption{The D-L view of the right component tree of subject $28$. The irregularities close to the root node indicate a possible starting point problem.}
\label{28right}%
\end{center}
\end{figure}

\item \textbf{Veins:} The tracking system used to obtain the data trees is intended to record only arteries in the brain. However, in some cases, veins that run very close to an artery are mistakenly identified as an extension of that artery. These veins are usually thicker than the parent artery, and show up as red leaf nodes on the visualization. An example of this kind of problem is displayed in Figure \ref{24back}.
\begin{figure}
[ptb]
\begin{center}
\includegraphics[
natheight=9.375400in,
natwidth=12.500000in,
height=4in,
width=5in
]%
{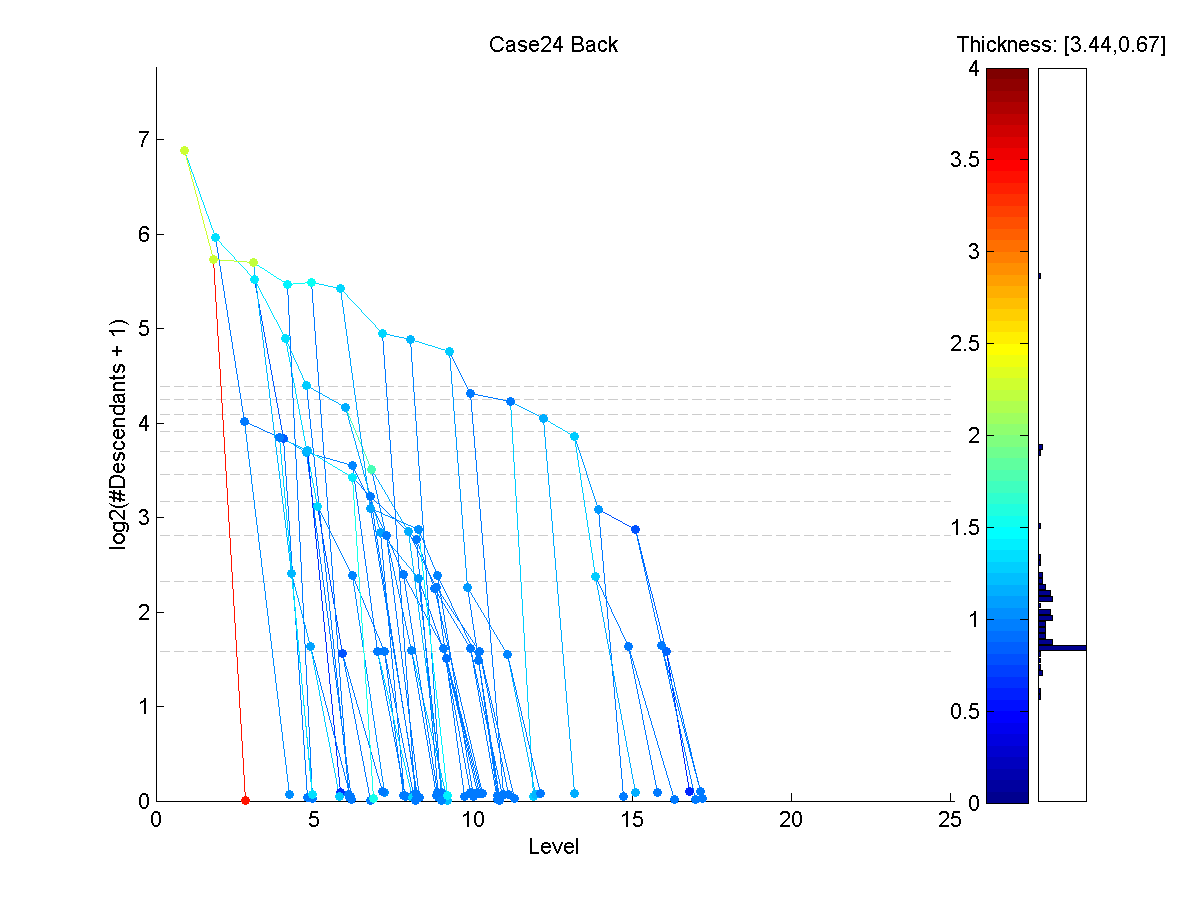}
\caption{The D-L view of the back component tree of subject $24$. The red leaf on the third level is identified as a possible vein.}
\label{24back}%
\end{center}
\end{figure}

\end{itemize}

\section{Solutions and Results}\label{Solution}
Through a careful inspection of the visual displays of all points in the data set, problematic instances are marked to be reviewed again. The numbers of instances marked for review for each sub-population are given in the top row of Table \ref{mark}.

\begin{table}[ptb]
\begin{center}
\begin{tabular}{|c|c|c|c| c|c|c|c| c|c|c|c|}
\hline
\multicolumn{4}{|c|}{Misconnection} & \multicolumn{4}{|c|}{Starting Point} & \multicolumn{4}{|c|}{Vein} \\
\hline
 B & L & R & F & B & L & R & F & B  & L  & R  & F    \\
 7 & 3 & 4 & 2 & 3 & 6 & 4 & 3 & 18 & 16 & 23 & 9 \\
 4 & 2 & 3 & 0 & 2 & 6 & 4 & 3 & 4  & 1  & 5  & 1 \\
\hline
\end{tabular}
\caption{Numbers of instances marked for reviewing (top row) and numbers of instances manually modified after raw data inspection (bottom row) for each kind of problem and for each tree location.}
\label{mark}
\end{center}
\end{table}

Note that the $4$ sub-populations contain a total of $420$ data trees. The process resulted in spotting $98$ of them ($24\%$) as potentially problematic instances that may significantly benefit from manual review of the $3-D$ raw data.

As a result of a careful study of the instances flagged, $7$ cases were identified as severely problematic (consistent with major errors in the scanning process), and were excluded from the data set. For the remaining cases, the bottom row of Table \ref{mark} shows the number of instances that were manually modified out of each problem group.

Of the remaining marked component trees, $35$ were selected for manual modification based on inspection of the raw data. The artifacts in the remaining $56$ instances were manually determined to be consistent with the natural noise.

While our visual diagnostic found many errors among the tree data, it is not perfect. In particular, one of the marked trees revealed an error different than flagged by the diagnostic.

Next, our visualization of the cleaned version of each of the above three example cases (with discrepancies) will be studied.

\begin{figure}
[ptb]
\begin{center}
\includegraphics[
natheight=9.375400in,
natwidth=12.500000in,
height=4in,
width=5in
]%
{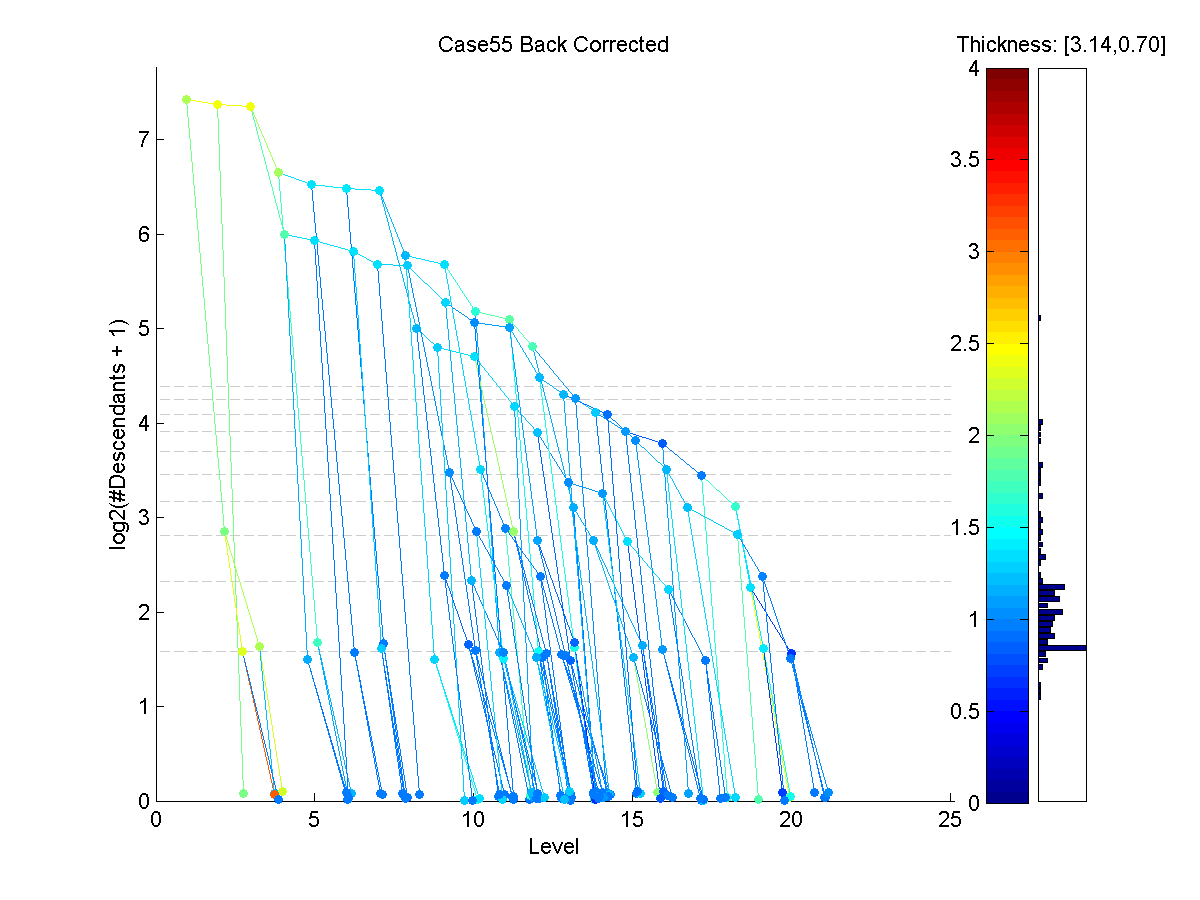}
\caption{The D-L view of the back component tree of subject 55 (Figure \ref{55back}) after correction. The misconnected subtree is deleted.}
\label{55backCorrect}%
\end{center}
\end{figure}

Figure \ref{55backCorrect} shows the corrected version of Case $55$, back component tree (from Figure \ref{55back}). Notice that the red/yellow subtree that was incorrectly attached to this tree has been removed.

\begin{figure}
[ptb]
\begin{center}
\includegraphics[
natheight=9.375400in,
natwidth=12.500000in,
height=4in,
width=5in
]%
{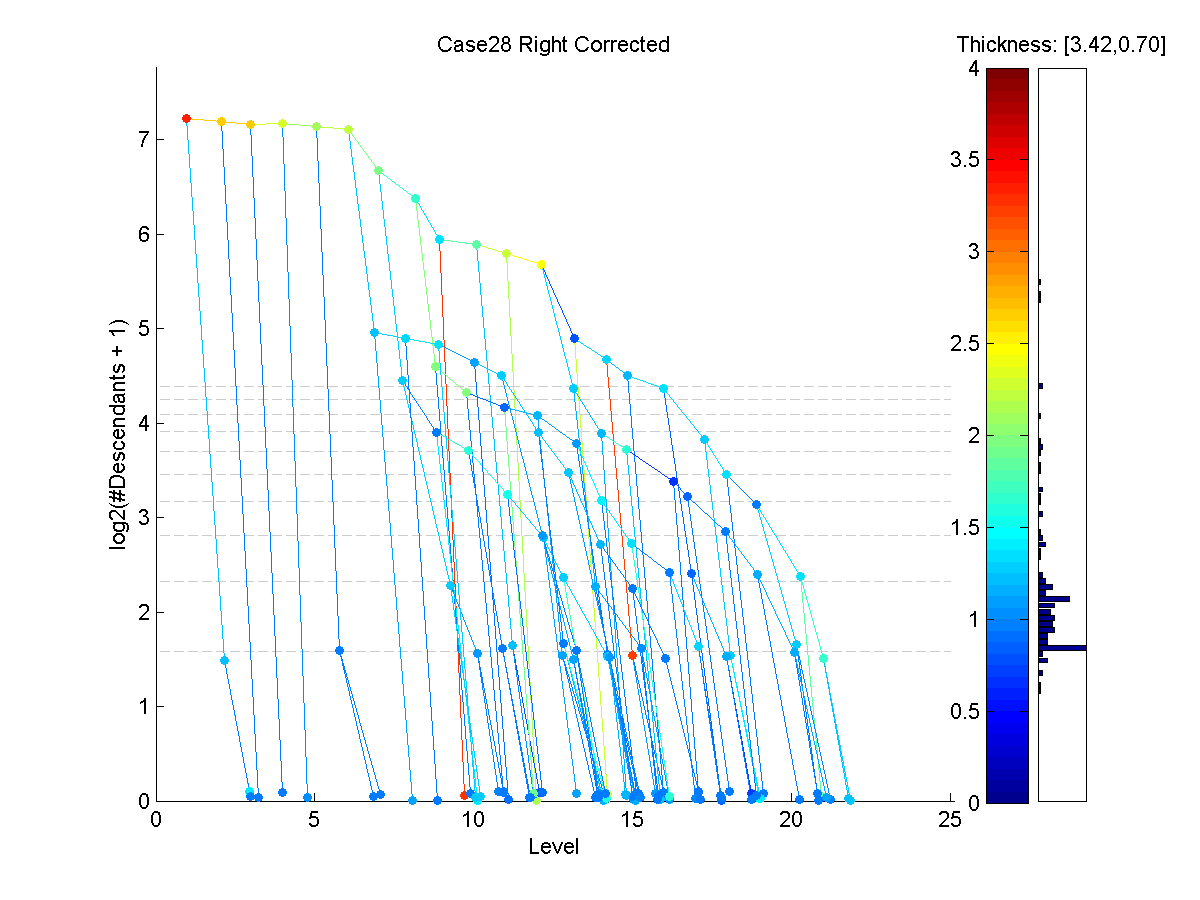}
\caption{The D-L view of the right component tree of subject 28 (Figure \ref{28right}), corrected by changing the cut-off point of the root vessel.}
\label{28rightCor}%
\end{center}
\end{figure}

Figure \ref{28rightCor} shows the result of fixing the starting point problem with the Case $28$ right component tree. Notice that the irregularity close to the root node (a series of dark red nodes) has been corrected by changing the cut-off level to a higher point.

\begin{figure}
[ptb]
\begin{center}
\includegraphics[
natheight=9.375400in,
natwidth=12.500000in,
height=4in,
width=5in
]%
{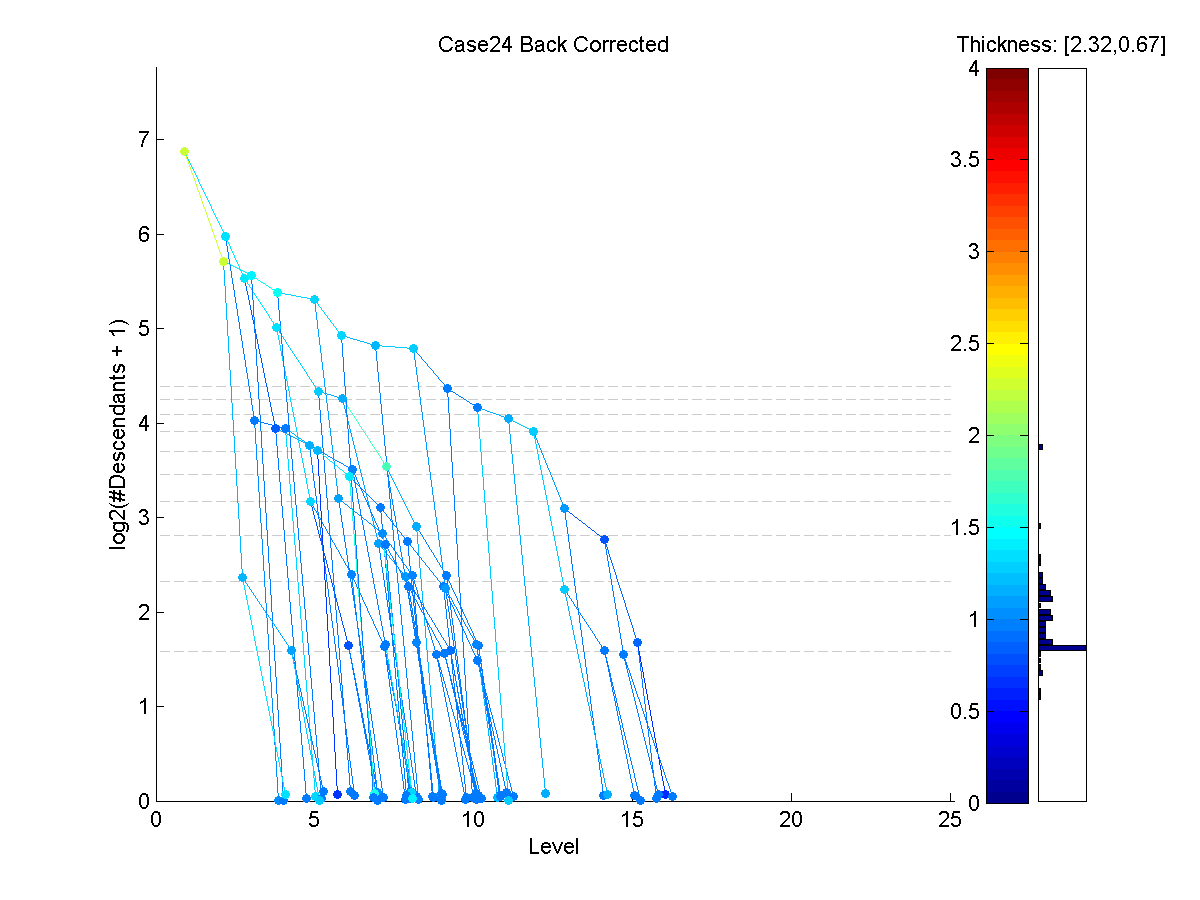}
\caption{The D-L view of the back component tree of subject $24$ (Figure \ref{24back}), corrected by the deletion of a vein.}
\label{24backCor}%
\end{center}
\end{figure}

Figure \ref{24backCor} shows that, after revision of subject $24$'s back component tree, the red leaf node seen in Figure \ref{24back} was indeed identified as a vein and was removed from the data tree.

All of these modifications in the data resulted in \texttt{Data Set $3$}. In particular, $7$ high-problem cases have been removed from \texttt{Data Set $2$} and the $35$ component trees have been corrected.

An important question regarding the clean up study is that: does the elimination of several problems improve the data set in terms of age effect? In other words, is the age effect more pronounced now or not? \cite{wang2007} and \cite{Aydin2009} have previously used a statistical analysis tool called \emph{tree-line analysis} to measure the effect of aging on brain vessel structure in a smaller data set (\texttt{Data Set $1$}). The same tool is used here to compare \texttt{Data Set $2$} and \texttt{$3$} in terms of age effect.

\begin{table}[ptb]
\begin{center}
\begin{tabular}{| l | c | c |}
\hline
        & Set 3     & Set 2  \\ \hline \hline
Back    & 0.0318    & 0.0715         \\
Front   & 0.0296    & 0.0436         \\
Right   & 0.0685    & 0.0743         \\
Left    & 0.0916    & 0.0493         \\ \hline
\end{tabular}
\caption{The slope $p$-values of first principal components obtained using the tree-line tool. Columns compare the \texttt{Data Sets $2$} and \texttt{$3$}. Smaller values point to greater significance level.}
\label{comppvaltab}
\end{center}
\end{table}

Table \ref{comppvaltab} summarizes the slope $p$-value comparisons of two data sets ($2\&3$) using the tree-line tool. It displays the the slope $p$-values obtained from first principal components against the age variable. Smaller $p$-values point to a greater statistical significance level, that is, a tighter relation between age and projected tree sizes. According to the findings of this analysis, the significance level increases ($p$-value decreases) in $3$ out of $4$ sub-populations with the cleaning. This means the age effect on brain vessel structure gets more pronounced after the cleaning process, in accordance with the expectations.

Although the cleaning process generated an improvement on age-projection size relation, the change is not dramatic. It is clear that the visualization process resulted in the correction of many instances and thus provided a more reliable data set. However, most of these corrections are local compared to the sizes of whole trees within the sub-populations. The effect of aging is related to the general trend of \emph{branchiness} within those trees, and this trend did not dramatically change with our cleaning: data trees with many branches still have a lot branches and vice versa.

The visualization method is not claimed to detect all possible problems that exist in the data sets. It relies on the irregular thickness differences of seemingly adjacent nodes. It is possible that there are still instances in the data set that contain for example misconnected vessels, but if those vessels wrongly connected have similar thicknesses, they will show up as normal branches on the visualization. However, there is no way of spotting those problems without using anatomical knowledge and the raw data, which would be an enormously time consuming task. The aim here is to provide a simple and quick new tool for finding problem instances.

In conclusion, the contribution of the visualization method is clear. $43\%$ of the instances that are marked for inspection revealed mistakes that it was appropriate to correct. As a result, the statistical significance of age effect on vessel structure improved, pointing to a less noisy data. The method provides an effective diagnostic for data clean up in similar populations of binary trees.

\bibliographystyle{abbrv}

\end{document}